\documentclass{llncs}
\usepackage[T1]{fontenc}
\usepackage{graphicx}
\usepackage{algorithm}
\usepackage[noend]{algorithmic}
\usepackage{amsfonts}
\usepackage{caption}

\newcommand\blfootnote[1]{%
  \begingroup
  \renewcommand\thefootnote{}\footnote{#1}%
  \addtocounter{footnote}{-1}%
  \endgroup
}

\begin{document}
\title{
    Towards A Complete Multi-Agent Pathfinding Algorithm For Large Agents
}
%
%\titlerunning{Abbreviated paper title}
% If the paper title is too long for the running head, you can set
% an abbreviated paper title here
%
\author{Stepan Dergachev\inst{1, 2}\orcidID{0000-0001-8858-2831} \and
Konstantin Yakovlev\inst{2, 1}\orcidID{0000-0002-4377-321X}}

\authorrunning{S. Dergachev et al.}
% First names are abbreviated in the running head.%

\institute{ 
    National Research University Higher School of Economics, Moscow, Russia \and
    Federal Research Center for Computer Science and Control of Russian Academy of Sciences, Moscow, Russia\\
    \email{sadergachev@edu.hse.ru, yakovlev@isa.ru}
    }
\maketitle              % typeset the header of the contribution

\begin{abstract}
Multi-agent pathfinding (MAPF) is a challenging problem which is hard to solve optimally even when simplifying assumptions are adopted, e.g. planar graphs (typically -- grids), discretized time, uniform duration of move and wait actions etc. On the other hand, MAPF under such restrictive assumptions (also known as the Classical MAPF) is equivalent to the so-called pebble motion problem for which non-optimal polynomial time algorithms do exist. Recently, a body of works emerged that investigated MAPF beyond the basic setting and, in particular, considered agents of arbitrary size and shape. Still, to the best of our knowledge no complete algorithms for such MAPF variant exists. In this work we attempt to narrow this gap by considering MAPF for large agents and suggesting how this problem can be reduced to pebble motion on (general) graphs. The crux of this reduction is the procedure that moves away the agents away from the edge which is needed to perform a move action of the current agent. We consider different variants of how this procedure can be implemented and present a variant of the pebble motion algorithm which incorporates this procedure. Unfortunately, the algorithm is still incomplete, but empirically we show that it is able to solve much more MAPF instances (under the strict time limit) with large agents on arbitrary non-planar graphs (roadmaps) compared to the state-of-the-art MAPF solver -- Continous Conflict-Based Search (CCBS). 

\blfootnote{This is a preprint of the paper accepted to MICAI'22}

\end{abstract}

\section{Introduction}
\label{sec:intro}

Multi-agent pathfinding (MAPF) is a challenging problem with topical applications in robotics, video games etc. There exist different ways to define the MAPF problem~\cite{stern2019multi} and approaches to solve it. On the one hand, optimal and bounded sub-potimal solvers exist for what is known as Classical MAPF, like the ones presented in~\cite{sharon2015conflict,boyarski2015icbs,barer2014suboptimal,sharon2012meta}. On the other hand fast prioritized planners without any completeness/optimality guarantees are widespread~\cite{cap2015prioritized,yakovlev2019prioritized}. Finally, complete, non-optimal algorithms exist, such as the ones described in~\cite{surynek2009novel,de2013push}, which borrow the solving techniques from the so-called  Pebble Motion on Graph (PMG) problem. Sill, all these algorithms do no consider the size/shape of the agents. Indeed, attempts to lift this restricting assumption are known~\cite{walker2018extended,andreychuk2021improving}, however the algorithms known so-far do not guarantee completeness. This works aims at drawing an attention to this gap and try narrowing it by adopting the PMG algorithms to the setting with the large agents, dubbed as MAPF-LA futher on.

More specifically, we focus on one of the routines, regularly needed, in solving MAPF-LA instances -- the one that makes it possible for one agent to traverse an edge without colliding to the other agents that prevent the transition due to their large bodies. We elaborate on how these agents can be safely moved away so the transition becomes valid. The suggested procedure was implemented and incorporated to the well-known \textsc{Push and Rotate} algorithm~\cite{de2013push}. Unfortunately, the current variant of this algorithm is still incomplete for MAPF-LA. However, as we show in our empirical evaluation, it outperforms the stat-of-the-art competitors, i.e. CCBS algorithms~\cite{andreychuk2021improving}, in terms of number of solved MAPF-LA instances under the strict time limit.

The rest of this paper is organized as follows. We considere the definitions of MAPF-LA in Section~\ref{sec:problem}. Section~\ref{sec:approach} describes a possible implementation of the procedure for moving away interfering agents, and also discuss cases when the proposed procedure is not sufficient to solve the planning problem. We report the results of of an experimental evaluation in Section~\ref{sec:experiment} and conclude in Section~\ref{sec:conclusion}.

\begin{figure}[tb]
    \centering
    \includegraphics[width=0.55\columnwidth]{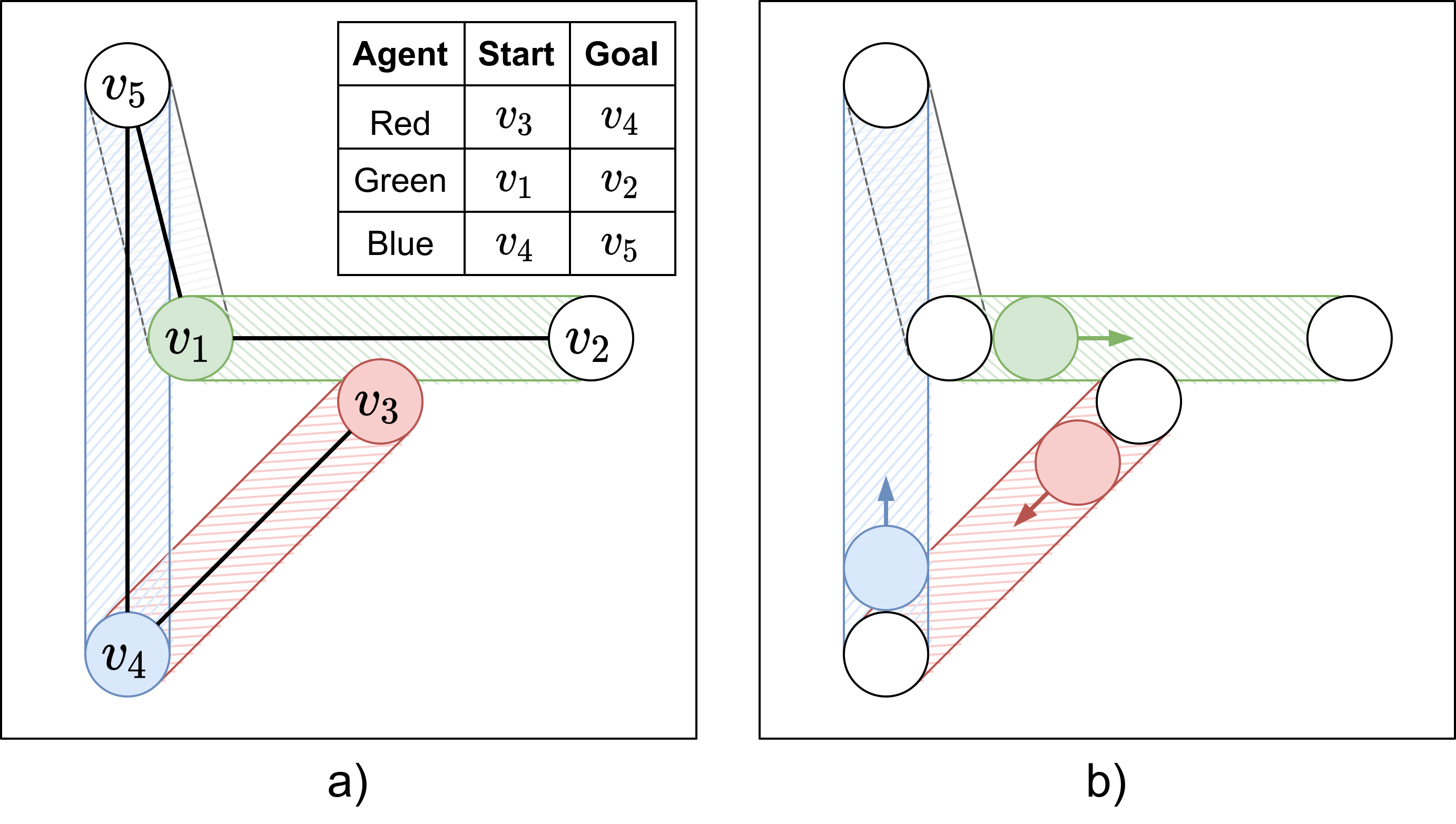}
    \caption{An example of the multi-agent path finding for large agents (MAPF-LA) problem instance. It can only be solved if the synchronous moves of the agents are allowed (as shown on the right). }
    \label{fig:example_synchronous}
\end{figure}

\section{Problem Statement}
\label{sec:problem}

Consider a tuple $(\mathcal{W}, G, r, K, start, goal)$, where $\mathcal{W} \subset \mathcal{R}^2$ is a metric workspace where $K$ agents are operating. Each agent is modeled as a disk of radius $r>0$. $G=(V, E)$ is a graph embedded in the workspace, i.e. each vertex $v\in V$ is associated with a point in $\mathcal{W}$. Edges correspond to the transitions between the locations. It is assumed that when moving along the edge the agent follows a straight-line segment connecting the corresponding vertices. $start: K \rightarrow V$ is a function that specifies the initial locations of the agents, $goal$ is the similar function specifying the target locations.

\begin{figure}[tb]
    \centering
    \includegraphics[width=0.6\columnwidth]{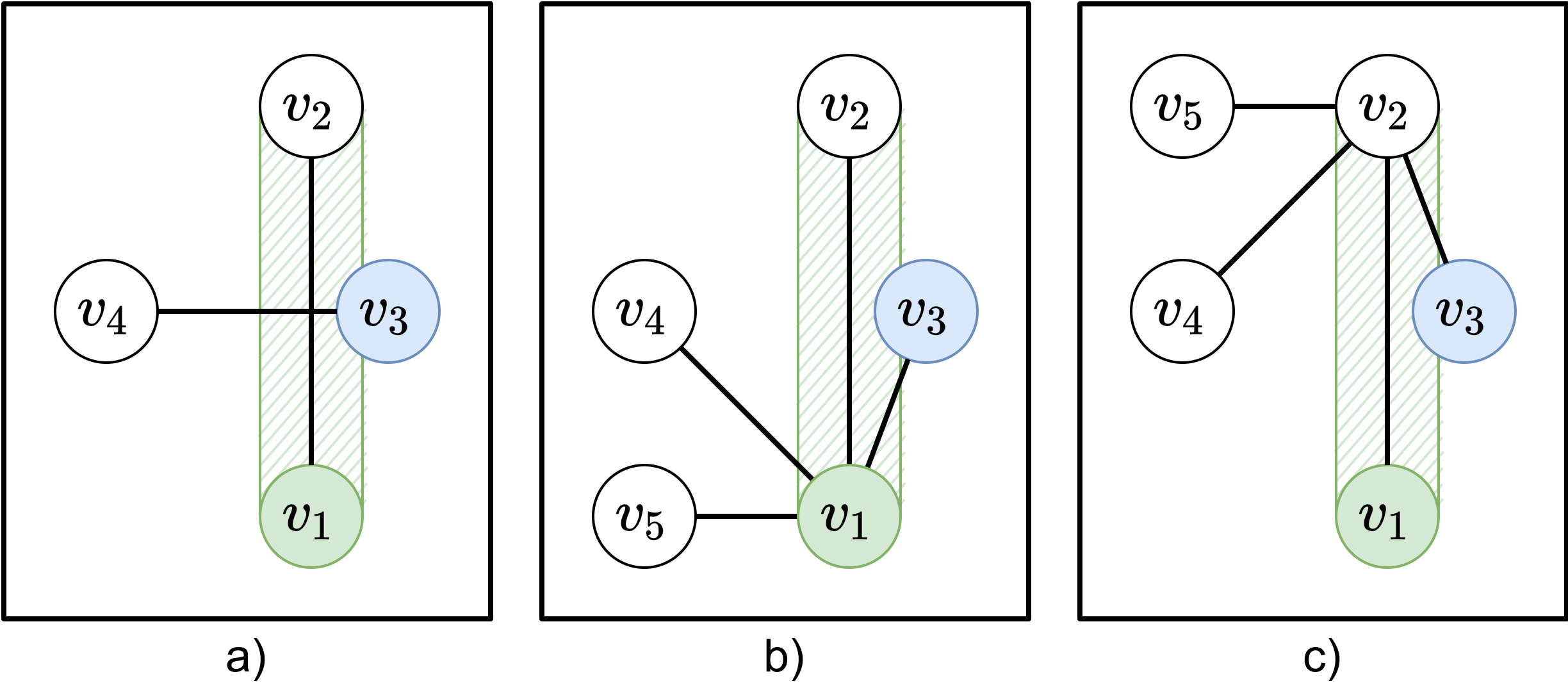}
    \caption{An illustration of three cases that can occur when agents should be moved away from the edge along which the current agent needs to transit}
    \label{fig:move_la_cases}
\end{figure}

A set of $K$ distinct graph vertices $S=(s_1, ..., s_K)$, $s_i \in V$, $\forall i,j: s_i \neq s_j$ forms a \emph{state}. Here $s_i$ is a position of the $i$th agent. A state is valid \emph{iff} $dist(s_i, s_j) \geq 2r$, where $dist$ stands for the Euclidean distance. In other words the state is valid if the bodies of the agents do not overlap. In this work we adopt an assumption that the distance between any two vertices of $G$ is greater than $2r$. This infers that any state, as defined above, is valid.

A \emph{transition} is formally a function $\pi(S, S') \rightarrow \{0, 1\}$, where $\pi=1$ stands for the valid transition and $\pi=0$ for the invalid one. Informally, transition corresponds to the movement of (some) agents between the locations in the workspace via the edges of the given graph. Conceptually, two possible assumptions regarding the transitions can be made:
\begin{itemize}
    \item general: synchronous moves of the agents are allowed, i.e. more than one agent can change its location as a result of the transition
    \item restrictive: only one agent can change its location, while the rest stay at the same vertices
\end{itemize}

In this work we follow the second assumption. In such case for a transition $\pi(S, S')$ to be valid the following conditions should be met. First, the moving agent $i$ may move only using one of the outgoing edges, i.e. $(s_i, s'_i) \in E$. Second, the move should be collision-free, i.e. $\forall j\neq i:  segdist(e, s_j) > 2r$, where $segdist$ is the distance between the segment, defined by the edge $e$, and the vertex $s_j$.

The problem of multi-agent pathfing for large agents (MAPF-LA) is now formulated as follows. Given $(\mathcal{W}, G, r, K, start, goal)$ find the sequence of states $(S_0, S_1, ..., S_n)$ s.t. $\forall s_i \in S_0: s_i = start(i)$, $\forall s_i \in S_n: s_i = goal(i)$, $\forall i=0,...n-1$ a valid transition $\pi(S_i, S_{i+1})$ exists. In other words, the problem is to find the sequence of moves for the agents that transfer them from their start locations to their goal locations, while avoiding the collisions.

\paragraph{Example.} An illustrative example of the considered problem is depicted in \figurename~\ref{fig:example_synchronous}. Here the graph $G$ consists of $5$ vertices and $4$ edges. The start and goal locations of the $3$ agents are specified on the left part of the figure. Basically, each agent has to transfer to the adjacent vertex. However, under the considered assumption that only one agent moves at a time, the instance is not solvable. Meanwhile, if synchronous moves are allowed the problem is trivially solved via a single transition in which the agents simultaneously move to the adjacent vertices (as shown on the right). This highlights how defining the transition influence the possibility to find a solution. This is similar to the pebble motion on graphs problem, in which different assumptions regarding cycle-moves and chains-moves can be adopted, see~\cite{yu2015pebble} for details.

\begin{figure}[tb]
    \centering
    \includegraphics[width=0.7\columnwidth]{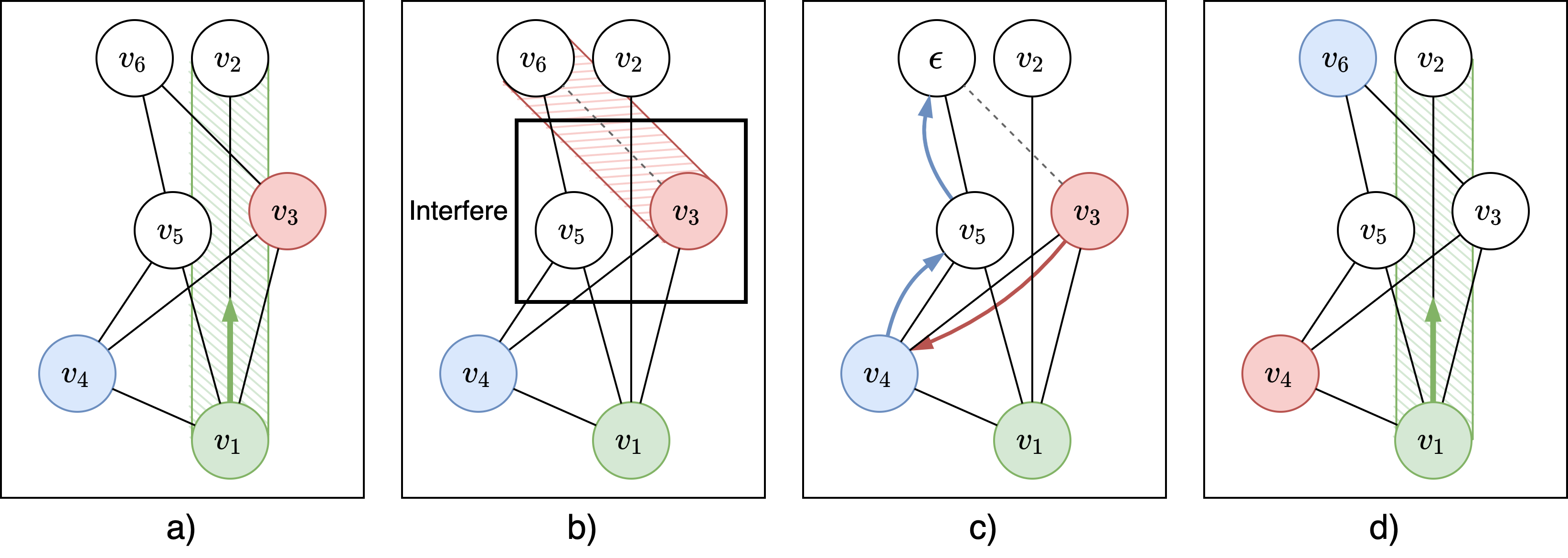}
    \caption{Illustration of execution of \textsc{PushToEmpty} procedure. (a) The green agent must move along the edge $(v_1, v_2)$, but the red agent interferes with it. (b) First of all, we mark all edges that are interfered with by the vertex $v_{2}$ as untraversable and find all interfering with $(v_1, v_2)$ vertices. (c) Single empty and non-interfering with $(v_1, v_2)$ vertex $v_6$ to which there is a path not through $v_1$ or $v_2$ remain on the graph. (d) After all, pushing the red and blue agents along the path between $v_3$ and $v_6$. Thus, the green agent can move along the edge $(v_1, v_2)$}
    \label{fig:push_to_empty}
\end{figure}

\section{Suggested Approach}
\label{sec:approach}
\subsection{Preliminaries}

The problem considered in this work, multi-agent pathfinding for large agents (MAPF-LA), is similar to the pebble motion on graphs (PMG) problem~\cite{yu2015pebble}. However the crucial differences exist, that prevent the straightforward application of the known methods to solve PMG for the considered formulation. Generally, two core differences between the MAPF-LA and PMG are as follows. First, in PMG every placement of pebbles (agents) on distinct graph vertices by definition form a valid state. In MAPF-LA this is not the case, however we adopted a (restrictive) assumption that the vertices of the given graph are located $2r$ distance units away from each other. Thus every placement of (large) agents on disjoint vertices also result in the valid state. 

Second, in PMG for a transition of one agent to be valid it is only required that the source vertex is free. In MAPF-LA this is not enough as the target vertex of the move can be free, but the moving agent, say $a$, may collide with the other agents while traversing an edge, as these agents stay too close to the edge itself. Such agents can be referred to as the \emph{interfering} agents w.r.t. to the given edge $e=(v_{from}, v_{to})$. Thus, in order to reduce (our formulation of) MAPF-LA to PMG the following problem should be solved. Given a valid state $S$ and an agent $a$ that needs to traverse the edge $e=(v_{from}, v_{to})$ find a sequence of the valid transitions $(\pi_0, ..., \pi_k, ..., \pi_{m})$ that results in the state $S'$, where the agent $a$ is located at $v_{to}$ and all other agents occupy the same vertices as in $S$. Here the transition $\pi_k$ corresponds to the move of the agent $a$, while the other transitions are moves of the other agents which are made, first, in order to remove the interfiring agents so the move through $e$ is possible, and, second, move all agents (except $a$) back their vertices.

\begin{figure}[t]
    \centering
    \includegraphics[width=0.5\columnwidth]{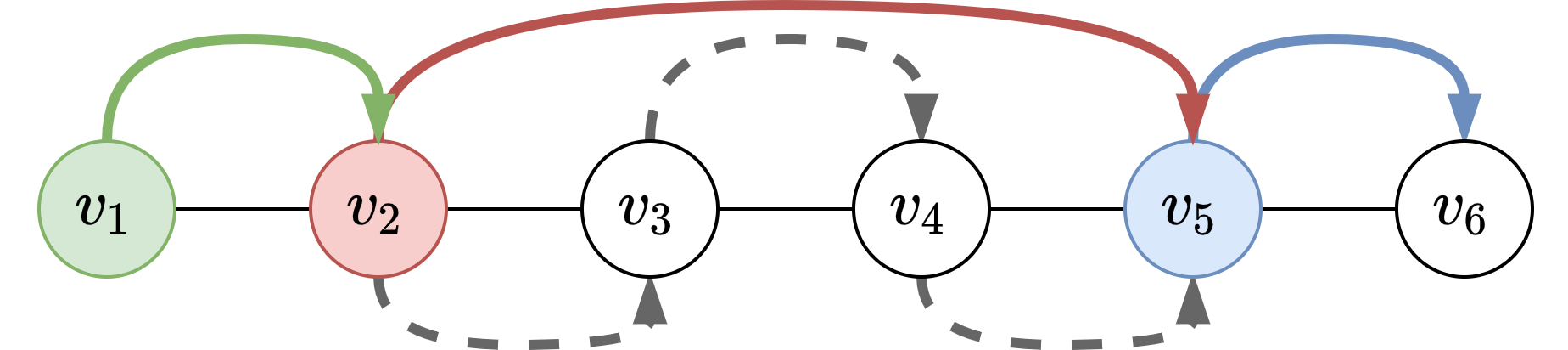}
    \caption{Illustration of execution of \textsc{PushAlongPath} procedure}
    \label{fig:push_along_path}
\end{figure}

Lets denote the procedure that solves the described problem as \textsc{move-la}. Next we elaborate on how this procedure can be constructed. Generally, after \textsc{move-la} is defined, one can use one of the PMG algorithms to solve MAPF-LA. E.g., in this work we use \textsc{Push and Rotate algorithm}~\cite{de2013push}, however other choices are possible.

\subsection{Moving Along An Edge}

A possible implementation of the \textsc{move-la} procedure consist of the three following steps. The first step is the sequential removal of every agent $a'$ from every vertex $v'$ that interfere with the move along $e$. This can be done via finding a path (for each interfering agent) to an unoccupied vertex and, then, sequentially moving the agent along this path (\textsc{push} operation). The second step is to move the $a$ along the $e$. The third step is to return every $a'$ to its initial vertex. However, this should be done in such a way that agent $a$ remains at $v_{to}$. We identify three different cases that may arise while removing an interfering agent $a'$ (Fig.\ref{fig:move_la_cases}). 

In the first case, a path to some empty vertex can be found for agent $a'$ that does not go through the vertices $v_{from}$ and $v_{to}$, as well as through the edges for which vertices $v_{to}$ are interfering. Thus, agent $a'$ can be pushed along such a path, and the resulting sequence of moves can be reversed at the end of the \textsc{move-la} operation to return agent $a'$ to the initial position. 
An example of such a case is shown in Fig.\ref{fig:move_la_cases}a, where the green agent needs to move along the $(v_1, v_2)$. To do this, it is enough to move the blue agent to vertex $v_5$, and after passing the green agent, return the blue one to $v_3$

\begin{figure}[t]
    \centering
    \includegraphics[width=0.7\columnwidth]{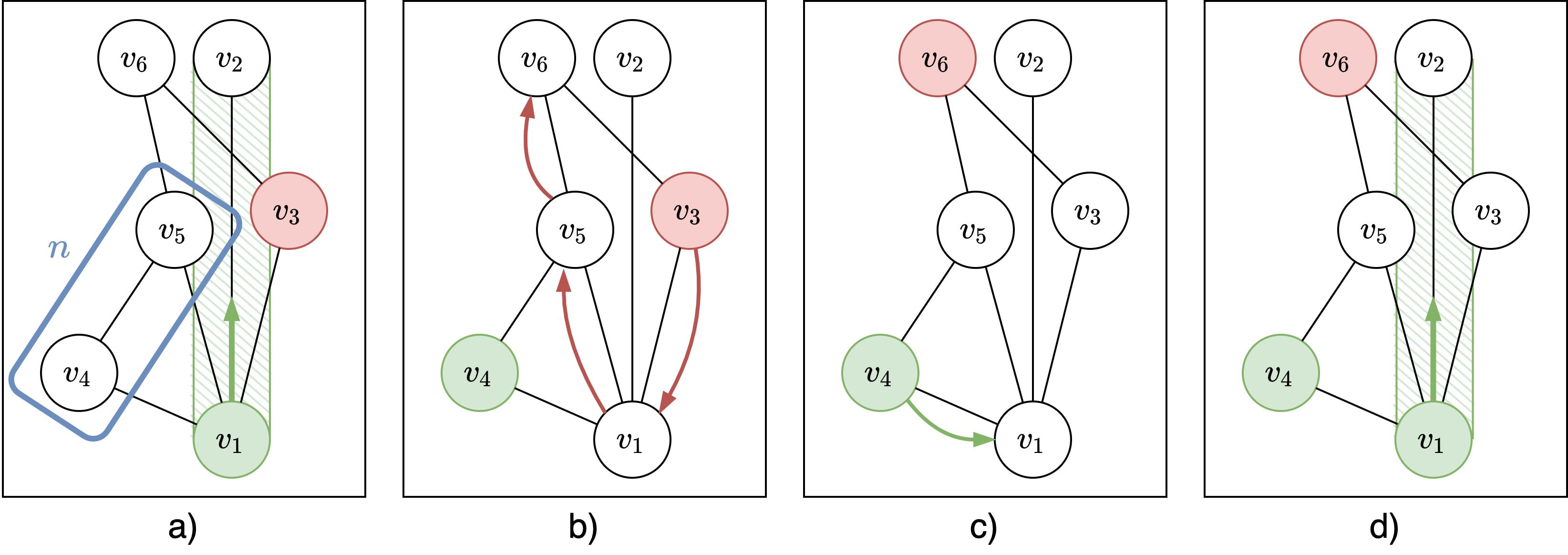}
    \caption{Illustration of execution of \textsc{PushThroughVFrom} procedure. (a) The green agent must move along the edge $(v_1, v_2)$, but the red agent on the $v_5$ vertex interferes with it. There are no paths between $v_5$ and non-interfering vertices, that do not pass through $v_1$. So, it is necessary to move the green agent at one of the neighboring vertices so that the red agent can pass (b) After that, it is necessary to move the interfering agent from its position using the \textsc{PushAlongPath} procedure (c) Next, the green agent can return to its original position $v_1$ (d) Thus, the green agent can move along the edge $(v_1, v_2)$}
    \label{fig:push_v_from}
\end{figure}

In the second case, the path of $a'$ can go through the vertex $v_{from}$, but not through the edges for which vertices $v_{to}$ are interfering. However, for this, it is necessary to first push agent $a$ from vertex $v_{from}$.Then we need to push $a'$ to some non-interfering vertex. After the pushing of agent $a'$ completes, agent $a$ must be returned. The sequence of moves obtained in this way can be also reversed to return all agents to their positions, except for some actions that must be ignored when carrying out the reverse operation. 

For the green agent on Fig.\ref{fig:move_la_cases}b, which illustrates the second case, to make the required move, it must first go to vertex $v_5$. After that, the blue agent will be able to go to vertex $v_4$, and the green agent can return to $v_1$ and move to $v_2$. Finally, the resulting solution can be reversed after deleting the moves of the green agent.

In the third case, to remove the agent from the interfering vertex, it is necessary to find a path through $v_{to}$ (and optional through $v_{from}$, but not through the edge $e$ itself), or through the edges, for which vertices $v_{to}$ (and optional $v_{from}$) are interfering. In this case, it is impossible to return the agents by a reversal, since by the time this operation is performed, the vertex $v_{to}$ will be occupied by agent $a$, which will lead to a collision. 

In the case on Fig.\ref{fig:move_la_cases}c, the blue agent can be moved to a non-interfering vertex only when passing through $v_2$. Thus, after the green agent moves to $v_2$, the blue agent can return to its original position $v_3$ only if the green agent misses it (e.g., when the blue agent leaves to $v_4$, the green agent must go to $v_5$ for the blue one to return to $v_3$).

We propose \textsc{PushToEmpty} and \textsc{PushThroughVFrom} procedures that can be used to solve the first and the second case respectively. Let's consider  these procedures in more detail.

\algsetup{linenosize=\scriptsize}
\begin{algorithm}[t!]
\scriptsize
\captionsetup{font=small}
\caption{\textsc{ReversableEdgeCleaning} procedure}
\label{alg:reversable_edge_cleaning}
\textbf{Input}: $G$ -- graph, $S$ -- current state, $e = (v_{from}, v_{to})$ -- edge to clear,\\
$U$ -- blocked vertices, $r$ -- agents radius\\
\textbf{Output}:\\
$\pi$ -- resulting solution,
$\pi_{rev}$ -- the sequence of valid moves, which returns interfering agents back\\

\begin{algorithmic}[1] 
    \STATE $\pi$ $\gets$ [], $a$ $\gets$ agent, that need to move by edge $e$, $NotCleared$ $\gets$ \{\}
    \STATE $I$ $\gets$ interfere vertices of $e$, $I_{f}$ $\gets$ unoccupied interfere vertices of $e$
    
    \FORALL{$v' \in I \backslash I_{free}$}
        \STATE $\pi'$ $\gets$ [], $S'$ $\gets$ $S$
    	\IF {PushToEmpty($G$, $S'$, $\pi'$, $v'$, $U \cup \{v_{to}\}$,  $e$, $I_{f}$, $r$)}
    	    \STATE $\pi$ $\gets$ $\pi + \pi'$, $\pi_{rev}$ $\gets$ $\pi_{rev} + \pi'$ except moves of $a$ 
            \STATE $S$ $\gets$ $S'$,  $I_{f}$ $\gets$ $I_{f} \cup \{v'\}$
        \ELSE
            \STATE $NotCleared$ $\gets$ $NotCleared \cup \{v'\}$
        \ENDIF
    \ENDFOR
    \FORALL{$v' \in NotCleared$}
        \STATE $a_{e}$ $\gets$ \{\}, $\pi'$ $\gets$ [], $S'$ $\gets$ $S$
        \STATE $\pi'' \gets$ PushThroughVfrom($G$, $S'$, $\pi'$, $v'$, $U$,  $e$, $I_{f}$, $r$)
        \IF{$\pi'' \neq false$}
            \STATE $\pi$ $\gets$ $\pi + \pi'$, $\pi_{rev}$ $\gets$ $\pi_{rev} + \pi''$ except moves of $a$ 
            \STATE $S$ $\gets$ $S'$, $I_{f}$ $\gets$ $I_{f} \cup \{v'\}$, $NotCleared$ $\gets$ $NotCleared \backslash \{v'\}$
        \ENDIF
    \ENDFOR
\STATE $U$ $\gets$ $U \backslash \{v_{to}\}$, $\pi_{rev}$ $\gets$ Reverse($\pi_{rev}$)

\IF{$NotCleared$ not empty}
    \RETURN false
\ENDIF
\RETURN $\pi$, $\pi_{rev}$

\end{algorithmic}
\end{algorithm}

\subsection{PushToEmpty}

The procedure \textsc{PushToEmpty} is designed to remove interfering agent $a'$ without affecting the vertex $v_{to}$ and without pushing agent $a'$ through $v_{from}$ (Fig.~\ref{fig:push_to_empty}a). First of all, mark all edges that are interfered by the vertex $v_{to}$ as untraversable (Fig.~\ref{fig:push_to_empty}b). This is necessary so that when the obtained solution is reversed, there are no conflicts with agent $a$ who passed using the considered edge $e$. After that, all possible options are considered to move agent $a'$ from the interfering vertex $v'$. For this, a set of empty, non-interfering with $e$ vertices is formed.  For each selected empty vertex $\epsilon$, an attempt is made to find a path, avoiding using $v_{from}$ and $v_{to}$ (Fig.~\ref{fig:push_to_empty}c) and pushing the agent $a'$ along this path (using \textsc{PushAlongPath} procedure). As a result, the agent $a'$ from the interfering vertex is moved so that agent $a$ can move along the edge $e$ (Fig.~\ref{fig:push_to_empty}d).

If the path to the vertex was found, but the \textsc{PushAlongPath} operation failed, then the edge, through which the operation \textsc{move-la} inside \textsc{PushAlongPath} could not be performed, is temporarily marked as untraversable and the path to $\epsilon$ has searched again. If the path to $\epsilon$ cannot be found, then the next empty vertex from the list is taken. A more detailed description of the procedure is provided in pseudocode in Appendix A.

\subsubsection{PushAlongPath}

Lets consider \textsc{PushAlongPath} procedure. It consists of the sequential moving of agents along path to empty vertex starting from the last agent in the path. An important feature of this operation is that when passing a path through empty vertices, they will remain free after the end of the operation. 

An illustration of this operation is shown in Figure~\ref{fig:push_along_path}. Agents (green, red and blue circles) must be pushed along path $v_1-v_6$. In addition to the last vertex $v_6$, the path also contains intermediate empty vertices $v_3$ and $v_4$. The operation starts by moving the blue agent. He goes to vertex $v_6$. All subsequent agents move to those vertices that were occupied by the previous ones (the red agent moves to the vertex $v_5$ passing through empty vertices, and the green agent moves to the vertex $v_2$). At the end of the operation, the first vertex of the path is freed because the last vertex of the path becomes occupied.

\begin{figure}[t]
    \centering
    \includegraphics[width=0.3\columnwidth]{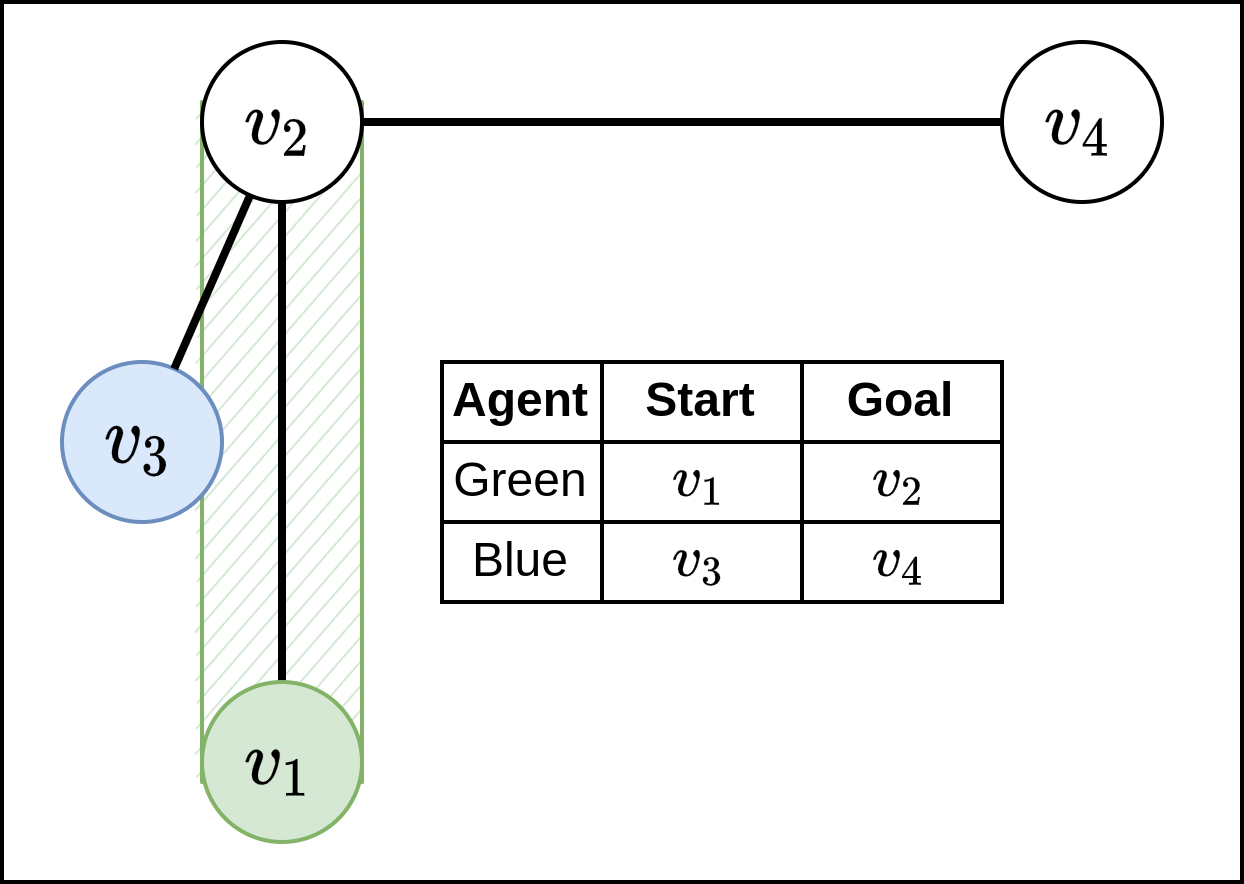}
    \caption{An example of a case where it is necessary to correctly determine the planning order, even if there is a complete procedure for moving an agent along an edge}
    \label{fig:order}
\end{figure}

\subsection{PushThroughVFrom}

To solve the second case of removing agent $a'$ from interfering vertex $v'$ we suggest the \textsc{PushThroughVfrom} procedure. It consists of two stages. At the first stage, agent $a$ moves away to one of the neighbouring vertices $n$. If $n$ is occupied, the operation of clearing the vertex is performed, similar to \textsc{PushToEmpty}. 

If the operation of clearing of $n$ was performed successfully, or if $n$ was initially not occupied, then the operation \textsc{move-la} from $v_{from}$ to $n$ is performed. Note that if the edge $e$ is marked as untraversable in the \textsc{move-la} operation, vertex $v_{to}$ can be involved in it, since the obtained actions $\pi''$ will not be included in the \textsc{reverse} operation when the interfering agents return to their vertices. In the case when at least one of the operations described above fails, then an attempt is made to move agent $a$ to another vertex $n$. 

After the passage through $v_{from}$ is cleared, an attempt is made to move agent $a'$ away from $v'$ also similarly to \textsc{PushToEmpty}. It is important to note that when creating the path of agent $a'$, it is necessary to block the vertices that are the current positions of the $\pi''$ participants. This is necessary to guarantee the return of agent $a$ to the vertex $v_{from}$ using the \textsc{reverse} operation. 

If it was possible to find a path to an empty vertex for agent $a'$, after which the \textsc{PushAlongPath} was successful, then the obtained solution is saved and supplemented with reverses actions $\pi''$ to return agent $a$ to vertex $v_{from}$. In addition, the solution is saved with the exclusion of action $\pi''$, which is necessary for the further return of agent $a'$ to vertex $v'$. A more detailed description of the procedure is also provided in pseudocode in Appendix A.

\begin{figure}[t]
    \centering
    \includegraphics[width=0.95\columnwidth]{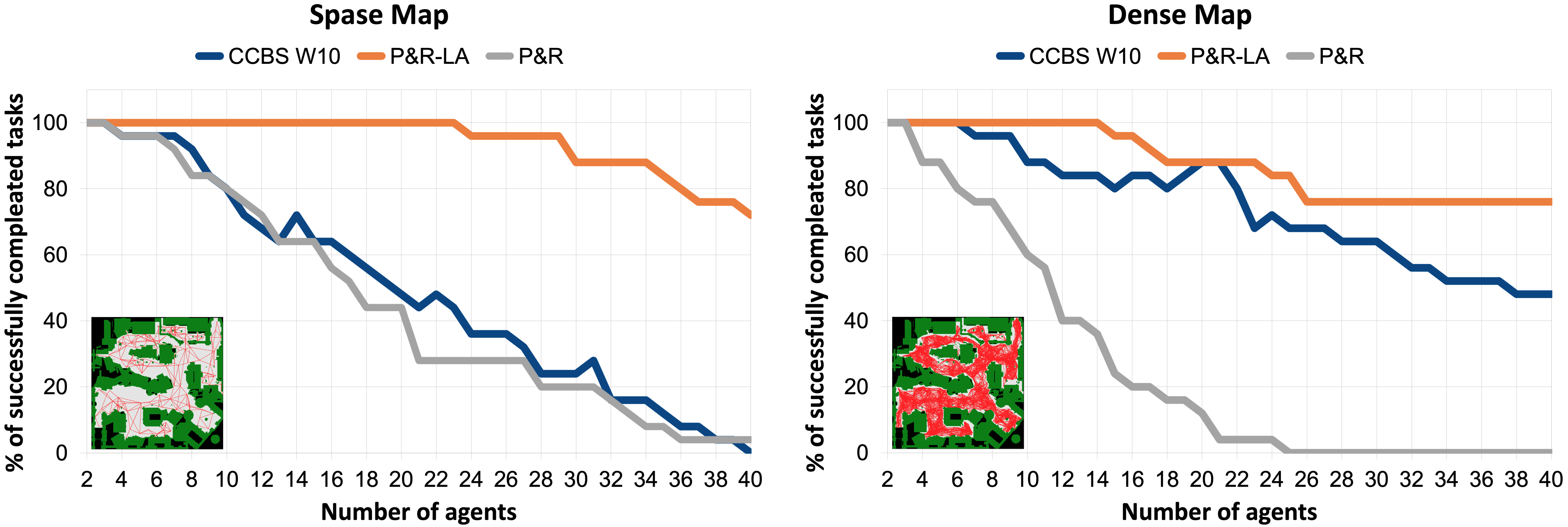}
    \caption{Success rates of the suboptimal version of \textsc{CCBS} (\textsc{CCBS W10}), the suggested approach (\textsc{P\&R-LA}) and  "naive" version of the \textsc{Push and Rotate} for large agents (\textsc{P\&R}) on two different roadmaps with varied number of agents}
    \label{fig:sr}
\end{figure}

\subsection{ReversableEdgeCleaning}

The sequential resolving algorithm of the first two cases can be combined into a single \textsc{ReversableEdgeCleaning} procedure (Algorithm \ref{alg:reversable_edge_cleaning}), the result of which is the removal of interfering agents possible for considered cases, as well as a sequence of moves that will return all pushed agents to their positions after agent $a$ passes along edge $e$. 

The proposed algorithm consists of two stages. At the first stage, the procedure \textsc{PushToEmpty} is launched for each occupied vertex that interferes with the edge $e$ (lines~6-12). If any of these vertices cannot be freed using procedure \textsc{PushToEmpty}, then they are stored in the $NotCleared$ set (lines~12). After that, the procedure \textsc{PushThroughVfrom} is executed for each element of the $NotCleared$ set (lines~13-19).

\subsection{Discussion}
It should be noted that at the moment the proposed algorithm is not complete for MAPF for large agents problem, since it does not take into account two significant points. 

The first point refers to the third case inside \textsc{move-la} procedure previously mentioned in the text. In this case, the path from interfering to the non-interfering vertex lies through the vertex $v_{to}$ or the edges that this vertex is interfering with. Then there is no possibility of returning the interfering agent to its original position by reversing its actions. Thus, agent $a$ must let agent $a'$ go to its initial position $v'$, or another path must be found that leads agent $a'$ to $v'$.

The second point is that may be the case that the interfering agent $a'$ cannot return to the vertex $v'$ at the end of \textsc{move-la} procedure (and \textsc{move-la} procedure fails), but still there exists a solution in which $a'$ can be moved from vertex $v'$ and not returned there. A trivial example of such a case is shown in Fig.~\ref{fig:order}. If the green agent has a higher priority during planning, then the blue agent cannot be pushed from vertex $v_3$ and returned there after moving the green agent. However, if planning starts with a blue agent, then a solution will be found. Thus, to construct a complete algorithm for MAPF-LA it is not enough to obtain a complete procedure for moving an agent along an edge. One of the possible solution to this problem is the choice of the correct planning order, as it is done in the \textsc{Push and Rotate} algorithm for solving not bi-connected instances.

\section{Experimental Evaluation}
\label{sec:experiment}
We incorporated the suggested procedures to the well-known MAPF algorithm \textsc{Push and Rotate}, so it can be used to solve MAPF-LA instances. We denote it as  \textsc{P\&R-LA}. As one of the baselines we have also implemented a ``naive'' version of the \textsc{Push and Rotate} for large agents that simply halts when trying to move an agent along an edge with other interfering agents present. This version is denoted as, simply, \textsc{P\&R}. Finally, the main baseline we were comparing with was the well-known \textsc{CCBS} algorithm~\cite{andreychuk2021improving} (we used the official implementation available on Github).

We evaluated planners on two different graphs (roadmaps) which were used in the CCBS paper: sparse and dense. The sparse roadmap contains 158 vertices and 349 edges, while the dense one -- 878 vertices and 7341 edges. Originally, these roadmaps were automatically generated based on the \texttt{den520d} map from the MovingAI MAPF benchmark set~\cite{stern2019multi}. An illustration of the roadmaps is shown in Figure \ref{fig:sr}.

For each roadmap 25 scenarios were created, each one involving with 40 non-overlapping start and goal vertices. In each scenario, the first $n$ start-goal pairs were selected, and then the evaluated algorithm was launched with a time limit of 30 seconds. In the experiment, the value of $n$ varied from 2 to 40. We also set the sub-otimality factor for CCBS to 10, which notably speeds up the search (this feature is not described in the original paper, however is supported in the authors' code).

The resultant success rates, i.e. the fractions of the solved instances per fixed number of agents, are presented in Fig.~\ref{fig:sr} (the higher – the better). As one can see, our modification of \textsc{Push and Rotate} outperforms the competitors, especially when the number of agents goes up. E.g. our algorithm managed to solve 80\% of tasks with 40 agents on the sparse roadmap, while the success rate of both competitors was close to zero. The reason why CCBS failed almost always is that the density of the agents, i.e. the ratio of agents' number to the number of graph vertices/edges is high and no easy solution is possible. This explains why CCBS copes better with the same number of agents on the dense roadmap -- here there are much more graph vertices/edges that the algorithm can use to find non-conflicting plans. Notably, the reasons why our algorithm failed to solve instances invovling large number of agents differ for different maps. In sparse roadmap it terminated without providing a solution, while in the dense one it was not able to finish within the time limit. Thus, we infer, that a more efficient implementation of our method can actually provide a better success rate on the dense map.

\section{Conclusion}
\label{sec:conclusion}

In this work, we have considered the problem of designing a complete algorithm for a challenging variant of the multi-agent pathfinding problem when the size of the agents has to be taken into account (MAPF-LA). We elaborate on one of the core procedures such algorithm should incorporate, i.e. the procedure that clears an edge to allow for one agent to use this edge for a safe transition. We embed our implementation of this procedure to the well-known MAPF algorithm \textsc{Push and Rotate} enabling it to solve MAPF-LA problems. The results of the empirical evaluation provided us with the clear evidence that the resultant algorithm is able to find solutions for non-trivial MAPF-LA instances involving dozens of agents under the strict time limits (while its competitors often fail to do so).

Indeed, the main direction of future research is to develop a provably complete algorithm for solving MAPF-LA, as the method proposed in this work is only a step towards this goal (as it does not guarantee completeness).

% ---- Bibliography ----
%
% BibTeX users should specify bibliography style 'splncs04'.
% References will then be sorted and formatted in the correct style.

\bibliographystyle{splncs04}
\bibliography{bibl}

\appendix

\algsetup{linenosize=\scriptsize}

\section{PushToEmpty and PushThroughVfrom procedures}

\begin{algorithm}[h!]
    \captionsetup{font=small}
    
    \caption{\textsc{PushToEmpty} procedure}
    \label{alg:push_to_empty}
    \scriptsize
    \textbf{Input}: $G$ -- roadmap, $S$ -- current state, $\pi$ -- solution, $U$ -- blocked vertices,\\
    $v'$ -- interfering vertex, $e = (v_{from}, v_{to})$ -- edge to clear, $r$ -- agents radius, $I_{f}$ -- unoccupied vertices interfering to $e$.
    \begin{algorithmic}[1] 
        \STATE $E$ $\gets$ edges for which vertex $v_{to}$ is interfering, $G'$ $\gets$ remove $E$ from $G$
        \FORALL{$\epsilon$ in EmptyVertices($G$, $S$) \textbackslash $I_{f}$}
            \WHILE{true}
                \STATE $p$ $\gets$ path from $v'$ to $\epsilon$ in $G'$ \textbackslash $U \cup \{v_{from}\}$)
                    \IF{$p$ = false}
                        \STATE \textbf{break while}
                    \ENDIF
                    \STATE $\pi'$ $\gets$ [], $S'$ $\gets$ $S$, $e_{f}$ $\gets$ PushAlongPath($G'$, $S'$, $\pi'$, $p$, $U$)
                    \IF{$e_{f}$ = false }
                        \STATE $\pi$ $\gets$ $\pi + \pi'$, $S$ $\gets$ $S'$, \textbf{return} true
                    \ELSE
                        \STATE Remove edge $e_{f}$ from $G'$ 
                    \ENDIF
            \ENDWHILE
        \ENDFOR
        \STATE \textbf{return} false
    \end{algorithmic}
    \end{algorithm}

\begin{algorithm}[h!]
\captionsetup{font=small}
\caption{\textsc{PushThroughVfrom} procedure}
\label{alg:push_through_v_from}
\scriptsize
\textbf{Input}: $G$ -- roadmap, $S$ -- current state, $\pi$ -- solution, $U$ -- blocked vertices,
$v'$ -- interfering vertex, $a'$ -- agent from $v'$, $e = (v_{from}, v_{to})$ -- edge to clear, $r$ -- agents radius, $I_{f}$ -- unoccupied vertices interfering to $e$.
\begin{algorithmic}[1] 
    \STATE $E$ $\gets$ edges for which vertex $v_{to}$ is interfering

    \FORALL{$n \in$ Neighbours($G$, $v_{from}$) \textbackslash $U \cup \{v'\}$}
        \STATE $G'$ $\gets$ remove edges $E$ from $G$, $e_{f} \gets$ false, $p \gets$ true
        \IF{$n$ is occupied}
            \FORALL{$\epsilon \in$ EmptyVertices($S$) \textbackslash ($I_{f} \cup \{v_{to}\})$}
                \STATE $e_{f} \gets$ false
                \WHILE{true}
                    \STATE $S'$ $\gets$ $S$, $\pi' \gets []$, $p$ $\gets$ path from $n$ to $\epsilon$ in $G'$ \textbackslash $U \cup \{v_{from}, v_{to}, v'\}$ 
                    \IF{$p$ = false}
                        \STATE \textbf{break while} 
                    \ENDIF
                    \STATE $e_{f}$ $\gets$ PushAlongPath($G'$, $S'$, $\pi'$, $p$, $U\cup\{v_{to}\}$)
                    \IF{$e_{f}$ = false}
                        \STATE \textbf{break for}
                    \ENDIF
                    \STATE Remove edge $e_{f}$ from $G'$ 
                \ENDWHILE
            \ENDFOR
        \ENDIF
        \IF{$p = $false \textbf{or} $e_{f} \neq $false}
            \STATE \textbf{continue} with next n
        \ENDIF
        
        \STATE $G''$ $\gets$ remove edge $e$ from $G$, $\pi'' \gets []$
        \IF{\textbf{not} move-la($G''$, $S'$, $\pi''$, $v_{from}$, $n$, $U$, $r$)}
            \STATE \textbf{continue} with next n
        \ENDIF
        
        \STATE $P \gets$ $I_{f} \cup $ vertices of $\pi''$ $\cup$ interfere vertices of $(v_{from}, n)$
        \STATE $U' \gets$ $U \cup \{v_{to}\} \cup $ current positions in S' of agents from $\pi''$  
        \FORALL{$\varepsilon \in$ EmptyVertices($S$) \textbackslash $P$}
            \STATE $e_{f} \gets false$
            \WHILE{True}
                \STATE $S''$ $\gets$ $S'$, $\pi''' \gets []$, $p$ $\gets$ path from $v'$ to $\varepsilon$ in $G'$ \textbackslash $U'$
                    \IF{$p$ = false}
                        \STATE \textbf{break while} 
                    \ENDIF
    
                    \STATE $e_{f}$ $\gets$ PushAlongPath($G'$, $S''$, $\pi'''$, $p$, $U\cup\{v_{to}\}$)
                    \IF{$e_{f}$ = false}
                        \STATE $\pi \gets \pi + \pi' + \pi'' + \pi'''$, remove all moves of $a'$ from $\pi''$
                        \STATE $\pi \gets \pi + reverse(\pi'')$, \textbf{return} $\pi' + \pi'''$
                    \ELSE
                        \STATE Remove edge $e_{f}$ from $G'$ 
                    \ENDIF
            \ENDWHILE
        \ENDFOR
    \ENDFOR
    \STATE \textbf{return} false
\end{algorithmic}
\end{algorithm}

\newpage

\end{document}